\documentclass{JHEP3}
\pdfoutput=1
\usepackage{graphicx,amsmath,amssymb}
\usepackage{bbm,bm,amsmath,amssymb}

\def\bg{\begin{eqnarray}}
\def\nd{\end{eqnarray}}

\title{Effective Theories as Truncated Trans-Series and Scale Separated Compactifications}
\author{Maxim Emelin
\vskip.03in
Department of Physics, McGill University \\
3600 rue University, Montr\'{e}al, Qu\'{e}bec, Canada H3A 2T8\\
{\tt maxim.emelin@mail.mcgill.ca}
}

\abstract{We study the possibility of realizing scale-separated type IIB Anti-de Sitter and de Sitter compactifications within a controlled effective field theory regime defined by low-energy and large (but scale-separated) compactification volume. The approach we use views effective theories as truncations of the full quantum equations of motion expanded in a trans-series around this asymptotic regime. By studying the scalings of all possible perturbative and non-perturbative corrections we identify the effects that have the right scaling to allow for the desired solutions. In the case of Anti-de Sitter, we find agreement with KKLT-type scenarios, and argue that non-perturbative brane-instantons wrapping four-cycles (or similarly scaling effects) are essentially the only ingredient that allows for scale separated solutions. We also comment on the relation of these results to the AdS swampland conjectures. For the de Sitter case we find that we are forced to introduce an infinite number of relatively unsuppressed corrections to the equations of motion, leading to a breakdown of effective theory. This suggests that if de Sitter vacua exist in the string landscape, they should not be thought of as residing within the same effective theory as the AdS or Minkowski compactifications, but rather as defining a separate asymptotic regime, presumably related to the others by a duality transformation.}

\begin{document}

\section{Introduction}

A complete theory of nature must be able to give a quantum mechanical account of gravitational interactions as well as reproduce the known low-energy physics of the Standard Model. The former requirement is predominantly a statement about the ultra-violet behavior of the theory, while the latter is a statement about the low-energy effective theory that arises in the infra-red. At first sight, satisfying these requirements seems like two distinct tasks, which can in principle be tackled separately. This is not the case. UV-completeness of the theory places tight constraints on the possible low energy theories that can arise. In the case of string theory, the set of such effective theories is called the \emph{string landscape}. The set of internally consistent effective theories that can not be realized in string theory are called the string \emph{swampland} \cite{vafa1}. Understanding the limits of these two sets is therefore crucial to any application of string theory to real world physics. Various conjectures have been proposed delineating the boundaries of the string landscape, known as the \emph{swampland criteria} \cite{swamplandreview}.

Perhaps the most important of these is the \emph{distance conjecture}. String theory has no fixed coupling constants. Every coupling in an effective theory is the expectation value of some degree of freedom in string theory. Likewise every expectation value in the EFT is secretly related to the mass or a coupling of some degrees of freedom in string theory. Pushing field values to extremes can therefore introduce new degrees of freedom or interactions invalidating the EFT description. The distance conjecture attempts to quantify this limitation of effective theories, stating that expectation values of fields must not exceed one Planck unit.

Another set of swampland conjectures has to do with compactifications to spaces with non-zero cosmological constant. These include several variants of the \emph{de Sitter conjecture} \cite{vafa-dS, Krishnan, dSrefine, noEI, TCC}, which effectively forbid long-lived phases of accelerated expansion. On the other hand, there are also conjectures relating to Anti-de Sitter space, specifically in the absence of supersymmetry and/or in cases of large scale-separation between the spacetime curvature and the Kaluza-Klein scale \cite{AdSdist, AdSnogo}.

Interestingly, these conjectures are in conflict with existing scenarios for achieving both a scale-separated AdS or dS solution, most notably the KKLT scenario \cite{KKLT} but also others \cite{BKQ, LVS}. These scenarios use EFT arguments and a vast literature is devoted to assessing their viability from a top-down perspective \cite{StarWars1, StarWars2} with arguments presented both for and against these proposals.

It is worth separating out two questions here. The first is whether dS or scale-separated AdS exist within string theory at all. The other is whether they exist within the regime of validity of any well-studied EFT limit of string theory. Note that the EFT-based proposals implicitly claim the answer is ``yes" to both questions, while the swampland conjectures claim the answer is ``no" to both. The top-down criticisms of the EFT scenarios, if correct, at most imply a negative answer to the latter question, without saying anything about the existence of these solutions in general. The present work also falls into this category in that it only aims at answering the latter question.

Many of the swampland criteria are statements about the effective action in the EFT. In other words they are off-shell statements. Meanwhile, worldsheet calculations in string theory only ever yield on-shell quantities. The off-shell action is then inferred from the on-shell observables. Of course in principle the quantum-corrected equations of motion contain the same information as the quantum-corrected effective action from which they are derived, but the power of off-shell field theory methods justifies thinking in terms of the effective action more than the associated EOM.

The situation changes however, once we start asking about whether a given solution can be realized after the incorporation of certain quantum corrections. This is an expicitely on-shell question and using the equations of motion may be more fruitful. 

In recent work \cite{ME0, ME, ME2, ME3}, the question of realizing de Sitter space was investigated from the equation of motion perspective. Working within the framework of supergravity with arbitrary quantum corrections, one can ask what quantum corrections would be required if they were to allow for a de Sitter solution. The result is that for time-independent compactifications there was no way to generate a de Sitter solution without switching on an infinite number of terms. This can be interpreted as a sign of a breakdown of the EFT description in similar spirit to the distance conjecture. Interestingly time-dependent compactifications were shown to avoid this problem \cite{ME2}, which resonates with the general picture provided by the swampland criteria although explicit connections remain to be drawn.

The goal of this paper is two-fold. One is to sharpen and expand the ideas behind the EOM-based approach of \cite{ME, ME2, ME3} and express them in a language of truncated trans-series. The other is to review and extend the analysis of those papers by studying the role of the warp factor and scale-separation in de Sitter as well as anti-de Sitter compactifications. 

The paper is organized as follows. In section \ref{EFT} we discuss the systematics of quantum corrections in effective theories where all couplings come from expectation values. We adopt the view that the quantum equations of motion in an asymptotic regime can be expanded in a resurgent trans-series, which neatly categorizes the various perturbative and non-perturbative corrections and define effective theories as truncations of that trans-series.

In section \ref{metric} we apply this approach to scale-separated compactifications to (A)dS and study the corrections that might allow for such compactifications to solve the truncated equations of motion without invalidating the truncation. An important aspect of our analysis is the existence of a family of corrections that don't scale with one of the expansion parameters that define our asymptotic regime. Along the way we find agreement with existing constructions of scale-separated AdS. Furthermore we argue that corrections that take the form of integrals over internal four-cycles are essentially the unique type of correction that allows for scale-separated AdS solutions. On the other hand, lifting to de Sitter appears to require bringing in the full family of non-scaling corrections, which destroys the EFT description.

We finish with a summary of our results and a discussion of their relationship to the swampland conjectures and KKLT-type constructions in section \ref{conclusion}.

\section{Equations of Motion, Quantum Corrections and Effective Theories} \label{EFT}

\subsection{Effective Theories as Trans-Series Truncations}

Consider a theory with degrees of freedom $\phi^i$ described by a full quantum effective action $\mathcal{S}[\phi_i ]$. The equations of motion coming from this action give the fully non-perturbative behavior of the on-shell observables. This theory can now be taken to an asymptotic regime defined by a set of parameters $g_k \to 0$ and we can attempt to expand the full action around that point in the following way:

\bg \label{qS}
\mathcal{S} =\mathcal{G} \sum_{\vec{n}, \vec{m}} e^{ - s_{\vec{n}} / e^{\vec{\alpha} \cdot \vec{n}  }}   ~  e^{ \vec{\alpha}\cdot \vec{m} } ~ S^{(\vec{n},\vec{m})}
\nd
where $\vec{\alpha}$ is a set of parameters $\alpha_k$ such that the perturbative series is expressed in powers of $g_k = e^{\alpha_k}$, $\mathcal{G}$ is the leading $g_k$ dependence, so that $S^{(0,0)}$ comes with no prefactor in the sum, and $\vec{n}, \vec{m}$ can then be considered to be strictly positive, $s_{\vec{n}}$ and $S^{(\vec{n},\vec{m})}$ are functionals of the $\phi_i$. The equations of motion can be then be expressed as

\bg \label{qEOM}
\sum_{\vec{m},\vec{n}} C_i^{(m,n)} &=& 0 \nonumber \\
C^{(\vec{n},\vec{m})}_i &=&   e^{ - s_{\vec{n}} / e^{\vec{\alpha} \cdot \vec{n}  }}   ~  e^{ \vec{\alpha}\cdot \vec{m} } ~ \left(  \frac{\delta S^{(\vec{n},\vec{m})}  }{\delta \phi_i} - e^{ - \vec{\alpha} \cdot \vec{n}  } \frac{\delta s_{\vec{n}}  }{\delta \phi_i} \right)
\nd

Both expression \eqref{qS} and \eqref{qEOM} deserve some commentary. Perturbative series are generally asymptotic, rather than convergent, so care is required when interpreting \eqref{qS} if we take the sums over all $\vec{m}$. On a more practical level, we typically start at the asymptotic regime and compute the corrections order by order, and would like a way to extract useful physics from the corrections that we can compute. Both of these concerns require us to establish when it is appropriate to truncate the series \eqref{qEOM} to some finite order.

For practical concerns, the answer is simple. An asymptotic expansion has a smallest term at some finite order, before the terms begin to grow combinatorially. A truncation to this order is called the \emph{optimal truncation} for the series and leaves a ``non-perturbative'' error, typically of order $\exp{(-1/g_k^l)}$, for some power $l$. If this error is larger than our desired precision for the effective theory, then we can not meaningfully truncate the series. Otherwise, we are free to truncate the series at the first term smaller than the desired precision and the resulting expression defines the effective theory.

If one has access to the full perturbative series, one can also try to \emph{Borel resum} it, which involves taking the inverse Laplace transform of each term, summing that series, which should now be convergent, and Laplace transforming the result. This is often impossible, due to poles appearing along the Laplace transform integration contour. Going around these poles introduces a ``non-perturbative'' imaginary ambiguity, also of order $\exp{(-1/g_k^l)}$. The power $l$ can be shown to be related to the asymptotic growth of the perturbative coefficients. 

It is possible to choose the coefficients $S^{(\vec{n},\vec{m})}$ so that the ambiguities arising in each sum over $\vec{m}$ for fixed $\vec{n}$ \emph{cancel} against another kind of ambiguity at different $\vec{n}$, rendering the entire series unambiguous and borel-resummable. In this case, we obtain what is known as a \emph{resurgent trans-series}. These objects have received some vigorous attention in various areas of physics in recent years (see \cite{transseries} for a review) and it has been conjectured that physical observables generally take this form.\footnote{A more general trans-series can involve nested exponentials and logarithms. Here we will assume the simplest case of the form \eqref{qS} and see how far we get.} We will assume that this is the case, so that \eqref{qS} and \eqref{qEOM} are mathematically rigorous representations of the fully quantum corrected action and equations of motion, respectively. This is not a particularly wild assumption, since pretty much all the functions that appear in physics admit a trans-series representation, which is unique once a choice of expansion parameters is fixed. Since the observables of the theory exist everywhere in parameter space, such a trans-series expansion around our asymptotic region will exist. Perturbative calculations in the asymptotic regime should then give precisely the $S^{(\vec{n},\vec{m})}$ coefficients.

For small values of $g_k$, the various terms in \eqref{qEOM} form a hierarchy and if we are happy with a certain degee of approximation $\epsilon$, we can truncate the series by eliminating all terms smaller than $\epsilon$.\footnote{Determining the ``size'' of a correction $C_i$ wold have to be done using the metric on field space obtained from the kinetic terms, i.e. $| C_i |^2 = \mathcal{K}^{ii} C_i C_i$.}  This will be our definition of an \emph{effective theory}. A similarly truncated \eqref{qS} will then be the effective action for that theory. This truncation only remains valid in a region of parameter space, which defines the regime of validity of the effective theory.  In contrast, the full untruncated trans-series expression remains well-defined for all values of the expansion parameters even well outside any regime where truncation is possible, and represents the full quantum effective action $\mathcal{S}$.

\subsection{The Breakdown of EFT} \label{dsp}

The Dine-Seiberg problem \cite{DS} is the statement that string theory vacua with fully stabilized moduli will generally reside in a regime where quantum corrections are comparable to tree level effects and one might then be worried that control over these corrections is lost. While this is a generic behavior one might worry about, it is of course not the case that every time a quantum correction is important for the behavior of a theory, that control over quantum corrections is lost. If this were the case, there could be no such thing as a one-loop effective action. Establishing under what conditions a particular ansatz forces us to leave the regime of control would therefore be useful.

Consider an expansion parameter $g$. If $g$ is an inverse mass scale, then the scaling of any term in the action or equations of motion is determined by the mass dimensions of the fields $\phi_i$ that enter it. If $g$ is a coupling that arises from the expectation values of the fields, this will also define a scaling of all operators $O(\phi_i)$ with respect to $g$. This is, of course, precisely what happens in string theory, where the only expansion parameters are the Planck (or string) scale and the expectation values of various fields (e.g. dilaton, moduli, fluxes). Once a background ansatz is chosen, the scalings of all the operators with respect to all the expansion parameters are fixed.

Let us take a such a theory in an asymptotic regime defined by $g_k \to 0$, for some parameters $g_k$. This immediately places certain constraints on the fields whose VEVs determine these parameters. However the remaining fields are at this point unconstrained. Here one of two things can happen. The first option is that all but finitely many operators in the theory have non-zero scaling with respect to at least one of the $g_k$. In this case, the operators that have zero scaling will become dominant in the asymptotic limit,\footnote{We assume there are no polar terms in the action, since we can always factor them out without affecting the equations of motion.} while the rest can be viewed as sub-leading corrections. In this case the remaining fields in the theory may remain unconstrained without endangering the validity of the EFT description, since the $g_k$ will always provide the necessary suppression of higher-order or non-perturbative terms. Thus a truncation to finitely many (although perhaps a very large number of) terms in the equations of motion will always be possible.

The other option is that the theory allows for infinitely many operators that have zero scaling with respect to $g_k$. These operators are disctinct, so they must be made of different powers\footnote{By ``powers" we will always mean all the various allowed contractions of any gauge indices the fields may carry, rather than simple multiplication.} of the various degrees of freedom of the theory. In this case, we must view the VEV of some of these fields as additional expansion parameters and include it in the definition of our asymptotic regime. 

This means that if we begin with an asymptotic regime defined by $g_k\to 0$ and find that we can construct an infinite family of operators $O_i$, containing different powers of a field $\psi$, such that 

\bg
O_i \sim \psi^{\Delta_i}
\nd

with $\Delta_i$ monotonically increasing with $i$, then $\langle \psi \rangle$ becomes an expansion parameter, and taking $\psi = \mathcal{O}(1)$ would turn on infinitely many terms in the full quantum equations of motion, preventing their truncation. This restricts the regime of validity of the EFT description beyond what is implied by the original $g_k \to 0$ definition. Put another way, the $g_k$ do not form a complete set of expansion parameters necessary to define a single EFT regime. The existence of the infinite family $O_i$ is then an indicator of this incompleteness.

The case that will be most relevant for our purposes is when a certain type of solution violates the equations of motion when $\langle \psi \rangle = 0$, but including, for example, $O_1$ in the equations of motion allows for the desired type of solution. Note that including $O_1$ at the equation of motion level, requires giving a VEV to $\psi$, which in turn, turns on \emph{all} the operators $O_i$. Preservation of the EFT descrition then requires that $\langle \psi \rangle$ provides the necessary suppression of the higher order operators, otherwise truncation of the EOM will become impossible.

Of course, which of these two scenarios occurs depends on the theory, in particular on its symmetries. The question boils down to whether defining an asymptotic regime for the theory requires restricting the VEVs of all the degrees of freedom or whether constraining only a subset is enough, with symmetries then forbidding the infinite family of non-scaling operators.

In the next section we will investigate the structure of allowed corrections in string theory, expanding in three quantities that define the limit of large volume type IIB compactifications at low energy. As a result, we will find ourselves in precisely the second scenario, where an infinite family of non-scaling corrections will be allowed, and control over them will have to be maintained by additionally imposing certain weak-field limits. We will find that AdS compactifications can be achieved within this limit, while their dS uplifts, will require large fields and will lead to precisely the breakdown described above.

\section{Scale-separated compactifications} \label{metric}

\subsection{The Ansatz} 

Let us consider the following type IIB background in Euclidean signature:

\bg \label{IIBmetric}
ds_{IIB}^2 &=&  - \frac{e^{\phi_B/2} }{H(y)^2} \frac{1}{ \Lambda x_0^2} \left( d x_0^2 + dx_1^2 + dx_2^2 + dz_1^2 \right) + e^{\phi_B/2} H(y)^2 ~\tilde{g}_{mn} dy^m dy^n ,
\nd
where $m=1,...,6$ and $\Lambda$ is a constant that sets the curvature scale of the non-compact space. Depending on the sign of $\Lambda$ this metric may be seen as a Wick rotation of either de Sitter or anti-de Sitter space, the minus sign in the first term is chosen so that positive $\Lambda$ corresponds to de Sitter space. Note that this means that $x_0$ is not the time coordinate in the AdS case, but the ``radial'' coordinate of the Poincar\'e patch. The metric is written in string frame, hence the dilaton factor. We will work with the M-theory dual of his metric, obtained via T-duality along the $z_1$ direction followed by uplift.

\bg \label{Mmetric}
ds_M^2 &=& H^{-8/3} \lambda^{-8/3} \left( d x_0^2 + dx_1^2 + dx_2^2  \right) + H^{4/3} \lambda^{-2/3} g_{mn} dy^m dy^n  \nonumber \\
&+& H^{4/3} \lambda^{4/3} \left( e^{-\phi_B} dz_1^2+ e^{\phi_B} dz_2^2 \right)
\nd 
where $z_2$ is M-theory circle coordinate and we have defined.

\bg
\lambda =  \sqrt{ - \Lambda  } x_0
\nd

Note that for $\Lambda > 0$, $\lambda$ becomes imaginary, but the appropriate Wick rotation to take our metric to de Sitter, makes it real in Lorenzian signature. The background will also involve fluxes, which we will require to preserve the isometries of the type IIB 4-dimensional spacetime. On the M-theory side this means that we can turn on the following global fluxes

\bg
G_{mnpa}, G_{0ijm} \nonumber\\
\nd
where the first components are dual to the usual type IIB bulk 3-form and 5-form fluxes we encounter in warped flux compactifications \cite{DRS, GKP}. Preserving the (A)dS isometries requires
\bg  \label{gextscale}
G_{0ijm} &\propto& \partial_m \left( \frac{\epsilon_{0ij} }{H^4 \lambda^4}   \right)  .
\nd
The $H$ and $\lambda$ dependence of the $G_{mnpa}$ components are determined by the flux equation of motion:

\bg
d\star G = G\wedge G + ... ~~,
\nd
where ``...'' denotes higher derivative terms.\footnote{e.g. the $X_8$ term coming from $\int C\wedge X_8$.} This implies that the $G_{mnpa}$ need to also have an intrinsic dependence, such that
\bg  \label{fluxscale}
G_{mnpa} G_{qrsb} \propto \lambda^0 ~ H^{4},
\nd
which we will split democratically among the various components, so that 
\bg \label{gintscale}
G_{mnpa} \propto H^2
\nd

As mentioned earlier, we will also allow for localized objects in our compactification, specifically D7-branes and O7-planes. On the M-theory side these are encoded in the internal metric $g_{mn}$, which differs from $\tilde{g}_{mn}$ that appears in \eqref{IIBmetric} by the addition of localized ``lump''-like features, such that

\bg
g_{mn} = \tilde{g}_{mn} + \sum_i g^{(loc)}_{(i)~mn}
\nd

where $i$ labels the various D7/O7 loci. The ``lump'' contributions vanish far from these loci. By continuity and diffeomorphism invariance, we expect them to be decaying functions of the square of the proper distance from the locus.\footnote{This can be explicitly checked to be the case for, say, a Taub-NUT space, which is dual to a D7 in flat space. While embedding into a general compactification manifold will certainly change the details of the localized metric, one can reasonably expect this feature to remain unchanged at least for some class of embeddings.}

\bg
g^{(loc)}_{mn} &=& g^{(loc)}(\rho^2)_{mn} \nonumber \\
\rho &=& M_p \int_\gamma  e
\nd
where $\gamma$ is the geodesic from the center of the ``lump'' to a given point and $e$ is the (warped) vielbein tangent to that geodesic. For simplicity, let us consider the constant coupling scenario, where all the type IIB D7-branes are on top of O7-planes such that their charges cancel  \cite{sen, KS}. This is simply to avoid having to include a variable axio-dilaton in our original ansatz. We are in principle free to move away from this point in moduli space. The whole M-theory metric thus interpolates between some ADE singularities at the D7/O7 loci and the type IIB compactification metric.

The localized ``lumps'' also support localized harmonic forms, which can be used to construct additional localized contributions to the flux.

\bg
G^{(loc)}_{mnpa} &=& F_{mn} \omega_{pa} 
\nd 
where $\omega_{pa}$ is a localized harmonic 2-form. The $F_{mn}$ are then dual to D7 worldvolume gauge field strengths. Again, by smoothness and diffeomorphism invariance we expect these forms to depend on the square of the proper distance from the center of their respective ``lump''.

\bg
\omega_{pa} &=& \omega(\rho^2)_{pa}   \nonumber \\
\rho &=& M_p \int_\gamma  e
\nd

\subsection{Scaling analysis}

In choosing our ansatz, we have done two separate but related things. We have introduced two functions, namely $\lambda(x_0)$ and $H(y)$. In doing so, we have also introduced two scales: the cosmological constant $\Lambda$ and the size of the internal manifold, which is given by the average magnitude the warp factor.The scale separation is measured by 
\bg
s = \Lambda \langle H^4 \rangle ,
\nd 
where angular brackets denote the average over the internal manifold on the IIB side. Large scale separation means small $s$. In a large volume compactification $\langle H^4 \rangle$ must be large in string units (but still small compared to the length scale at which we're studying the theory), so large scale separation and weak string coupling means $\Lambda \ll 1$.

Let us emphasize that the decision to study the M-theory description of the ansatz is only for convenience of bookkeeping. We are still interested in the regime of small $\lambda$ (corresponding to late times in dS or the region near the boundary in AdS). This is precisely the regime where the M-theory torus becomes small and the system is described by type IIB string theory. 

When looking for solutions in a given duality frame, one may worry that the corrections one ignores become large as one moves to another duality frame. This is true, when one is looking to truncate the off-shell effective actions in both descriptions. In particular if we find a solution to the truncated quantum EOM in type IIB, it may not be a solution to the truncated quantum EOM in M-theory, since the off-shell truncations are not the same.

This is not what we will be doing. Instead, we will be plugging the ansatz into the full quantum equations of motion, expressed as a trans-series, and systematically studying the scalings of all the quantum terms one could write down. These scalings will then dictate which terms become large or small in various limits, including our small-torus limit, and since we're still dealing with the full quantum effective action, this will be independent of the choice of duality frame. Once the scalings are determined, we can see if a truncation to an effective field theory description is possible.

We are interested in maintaining control in the type IIB limit with large compactification volume, which means we should be expanding in $\lambda$, $H^{-1}$ and $M_p^{-1}$. The limit when these are small, gives the small-torus, large volume, low-energy limit, i.e. type IIB supergravity (plus corrections). The small-torus limit may appear dangerous from the M-theory frame, but the corrections that become large in this limit come from M2-branes wrapping the torus, which are dual to momentum modes along $z_1$ on the IIB side. These modes are precisely part of our 4D EFT and should indeed be present. This is essentially the same as F-theory, except that that rather than taking the small torus limit manually, we run into that limit dynamically at small $\lambda$.\footnote{i.e. late times in dS or near the boundary in AdS.}

Our goal now is to determine the scaling of various terms in the EOM with respect to $\lambda$ and $H$ as well as the Planck mass, $M_p$. For terms to be able to cancel each other in the equations of motion, they need to have the same scaling with respect to these quantities. This is particularly true of $\lambda$ and $H$ scalings, since these are coordinate-dependent functions, and a mismatch between different orders would mean the terms don't even have the same functional form.

It will be convenient to study the equations of motion with one index raised, i.e.

\bg
G^M_{~~N} - T^M_{~~N} = g^{MP} \frac{\delta S}{\delta g^{PN} }
\nd
This way the scalings of each term will be the same as the scalings of the term in the action from which they originated, whereas keeping both indices lowered would shift the scalings of different components by different amounts depending on the subspace. For this reason, we will often use the scalings of the action (where all indices are contracted) as a proxy for the scalings of the equations of motion, although the exact form of the terms will differ. We now study the $\lambda, H$ and $M_p$ scaling of the EOM terms coming from the bulk fields, localized fields as well as non-local effects.

\subsection{Local Bulk terms}

A general non-topological term in the action is made up of Riemann tensors and fluxes, with all indices contracted using either inverse metrics of contravariant levi-civita tensors. Note that due to the block-diagonal form of our metric ansatz, the indices of any metric factors will be raised with the same subspace inverse metric. This leads to some convenient cancellations when calculating the scalings coming from curvature factors.

The Riemann tensor with two upper indices has the schematic form

\bg
\mathcal{R} = g^{-1} \partial g^{-1} \partial g + g^{-1} (g^{-1} \partial g)^2
\nd

The warped product structure of our ansatz guarantees that at the two-derivative level the curvature components that contribute to the EOM will have indices on the derivatives that belong to the same subspace.\footnote{There do exist non-vanishing Riemann curvature components with one $x_0$ and one $y$ derivative, which can contribute at higher derivative level, provided they come in pairs. The scaling of these terms will be the same as that of a product of two curvatures, one with two $x_0$ derivatives, the other with two $y_m$ derivatives.} The overall scaling ends up simply being that of the inverse metric from the subspace along which the derivatives act. This gives two types of curvature terms, those with two $x_0$ derivatives and those with two $y_m$ derivatives.

First note that all the $x_0$ dependences in our ansatz are power laws. For any quantity $X$ that depends polynomially on $x_0$, we have
\bg \label{lambdaExtra}
\partial_0 X \propto \frac{1}{x_0} X = \sqrt{-\Lambda}~ \lambda^{-1} X
\nd
The $0$ index will eventually be contracted with an inverse metric, so the overall effect of each derivative on the scaling is to introduce a factor of $\sqrt{-\Lambda}\lambda^{1/3} H^{4/3} M_p^{-1}$.

This means that curvatures with $x_0$ derivatives will contribute to the scaling as

\bg
R_{(ext)} \sim -\Lambda \lambda^{2/3} H^{8/3} M_p^{-2}
\nd
The subscript $(ext)$ denotes that the derivatives act along the external space, i.e. $x_0$.

The $y$ dependence appears in the unwarped 6d metric and possibly in the torus metric if we had chosen a non-trivial dilaton profile. Outside of that, all other $y$-dependence is encoded in the $H$-dependence, which is also always polynomial. This means we have

\bg
\partial_m g_{MN} \propto \frac{\partial_m H}{H} g_{MN} + ...
\nd
where ``..." denotes terms where the derivative hits the unwarped metric and thus doesn't alter the $H$ scaling. By diffeomorphism invariance, we expect that the combinations of derivatives hitting the metric to assemble into powers of $\Box H$ or $| \nabla H |^2$ and other covariant combinations. So the overall effect of these derivatives will be to give additional factors of 

\bg
\frac{\Box H}{H} \quad \text{or} \quad \frac{|\nabla H|^2}{H^2} \quad \text{etc.}
\nd
These quantities can be big or small in different locations on the internal manifold. Einstein's equations will then imply conditions on either the $y$-dependence of the flux terms or on the functional form of $H(y)$ itself. Factoring out these quantities, the remaining $H$-scaling is the same as that of the term inside the derivative. Thus, up to a factor of some function $f(\frac{\nabla H}{H} ,\frac{\Box H}{H} , ... )$, the curvatures with $y_m$ derivatives scale as

\bg
R_{(int)} \sim \lambda^{2/3} H^{-4/3} M_p^{-2}
\nd 
with the subscript $(int)$ denoting that the derivatives act along the internal space.

The fluxes also fall into two categories. There are the $G_{0ijm}$ and $G_{mnpa}$ components, which we will collectively denote by  $G_{(ext)}$ and $G_{(int)}$, respectively. They have some intrinsic $\lambda$ and $H$ dependence given by \eqref{gextscale} and \eqref{gintscale}. Since contracting a pair of indices, multiplies by one factor of the inverse metric, the total contribution to the scaling from each factor of the flux is simply determined by multipling by the square root of the inverse metric scaling for each index. We thus get

\bg
G_{(ext)} &\sim& ( \lambda^{-4} H^{-4} M_p^{-1}) (\lambda^{12/3 + 1/3} H^{12/3 - 2/3}) =\lambda^{1/3} H^{-2/3} M_p^{-1} \nonumber \\
G_{(int)} &\sim&  ( \lambda^{0} H^{2} M_p^{-1}) (\lambda^{3/3 - 2/3} H^{-6/3 - 2/3}) = \lambda^{1/3} H^{-2/3} M_p^{-1}
\nd

Finally we can also have additional derivatives acting on our fields, to give higher derivative corrections. They contribute an extra $M_p^{-1}$ to the scaling as well as another square root of the inverse metric that is used to contract them. In other words their contribution to the scaling of a term in the action is $\lambda^{1/3}H^{-2/3} M_p^{-1}$ for $\nabla_m$ and $\lambda^{1/3} ~H^{4/3} ~M_p^{-1}$ for $\nabla_0$. Table \ref{globalscales} summarizes the contributions to the scaling from the bulk fields.

\begin{table}[h!]
\centering
\begin{tabular}{||c |c ||} 
 \hline
 Field & scaling \\ [0.5ex] 
 \hline\hline
$R_{(ext)}$ &  $-\Lambda~ \lambda^{2/3} ~H^{8/3} ~M_p^{-2}$\\
 $R_{(int)}$ & $~~~\lambda^{2/3} ~H^{-4/3} ~M_p^{-2}$ \\ 
 $G_{(ext)}$ & $~~~\lambda^{1/3} ~H^{-2/3} ~M_p^{-1} $ \\
 $G_{(int)}$ & $~~~\lambda^{1/3} ~H^{-2/3} ~M_p^{-1} $\\
 $\nabla_m$ & $~~~\lambda^{1/3} ~H^{-2/3} ~M_p^{-1}$  \\
 $\nabla_0$ & $~~~\lambda^{1/3} ~H^{4/3} ~M_p^{-1}$  \\ [1ex] 
 \hline
\end{tabular}
\caption{Scaling contributions from bulk fields after complete contraction. The derivatives in the last two lines are assumed to be acting on bulk fields.}
\label{globalscales}
\end{table}

These results reveal our primary challenge in constructing scale-separated solutions with non-zero cosmological constant: $R_{(ext)}$ does not have the same $H$-scaling as any of the other 2-derivative terms. The other terms all have the same scaling and can therefore conceivably cancel against each other in the equations of motion, but not the terms coming from $R_{(ext)}$. This is a manifestation of the pure supergravity no-go theorems for de Sitter space \cite{MNnogo} and scale-separated anti-de Sitter space \cite{AdSnogo}. We must look for other ingredients beyond the 2-derivative bulk fields. 

In the following rest of this section we will examine two types of additional ingredients which may be present, namely localized terms and non-local corrections. The former come from localized objects in the compactification, such as branes and orientifold planes. These provide qualitatively different contributions to the equations of motion, however due to their localized nature, they can not help solve the equations of motion everywhere in the bulk. On the other hand, non-local corrections in the form of integrals over subspaces can be used, among other things, to capture non-perturbative effects such as brane-instantons. These corrections are present everywhere in the bulk and can be interpreted as large virtual fluctuations of (partially) localized objects, which allows us to dress the non-perturbative terms with additional bulk as well as localized terms. By studying these terms we will find the form of the corrections necessary to cancel $R_{(ext)}$ in the equations of motion, but we will also find a family of operators that do not scale with $\lambda$. As discussed previously, this raises the possibility of the breakdown of effective field theory, if the other expansion parameters do not provide adequate suppression, and we will discuss the conditions under which it occurs.

\subsection{Localized terms and a family of non-scaling operators}
\label{locterm}

As discussed previously, the presence of localized objects in our compactification ansatz, specifically D7 branes and O7 planes introduces localized modifications to the compactification manifold metric. Near one of these localized objects, the metric and inverse metric can be written as

\bg
g_{mn} = \tilde{g}_{mn} + h_{mn}  \nonumber \\
g^{mn} = \tilde{g}^{mn} + \bar{h}^{mn} ,
\nd
where $h_{mn}$ and $\bar{h}^{mn}$ satisfy

\bg
h^{m}_{n} + \bar{h}^{m}_{n} + \bar{h}^{mp} h_{pn} = 0,
\nd
with indices are raised and lowered using only $\tilde{g}$. This condition ensures hat $g^{mp}g_{pn} = \delta^m_n$. Note that $h$ is not a small perturbation, so we can not linearize in it. We also make no assumptions about its tracelessness etc. Indeed, near the localized objects, we expect $h$ to dominate the behavior of the metric, even the trace part, giving the ``near-brane'' geometry.

The metric along the $(a,b)$ directions also changes and deviates from that of the flat torus:

\bg
g_{ab} = \tilde{g}_{ab} + h_{ab}
\nd

which captures the non-trivial fibration of the M-theory circle in that region and allows for non-trivial topological charges. These are essential to cancelling the total charge from any internal fluxes and M2-branes present.

Accompanying this deformation is the presence of a normalizable 2-form $\omega$, with legs transverse to the extended object, which can be used to construct localized fluxes, which are dual to worldvolume fluxes on the D7 branes in type IIB.

\bg
G_{mnpa}^{(loc)} = F_{mn} \omega_{pa},
\nd

where the $mn$ indices are along the worldvolume.

The exact form of the near-brane geometry and the 2-form will generally be some complicated deformation of the ``lump'' solutions corresponding to the uplifts of the appropriate branes in flat space. The only assumptions we will make regarding these solutions are that $h_{mn}, h_{ab}$ have the same intrinsic scaling with $\lambda$ and $H$ as $\tilde{g}_{mn} as the ``unperturbed" metric$\footnote{or equivalently that $h^m_n$ and $\bar{h}^m_n$ have no scaling} and that $h_{mn}, h_{ab}$ and $\omega_{pa}$ depend on the \emph{square} of the proper distance from the lump center.

The proper distance from the center takes the form

\bg
\rho &=& M_p \int_\gamma  e,
\nd
 where $\gamma$ is the geodesic from the center to our point and $e$ is the vielbein tangent to it. For any function of $\rho^2$ we will then have
 
 \bg \label{dperp}
 \partial_{y_\perp} f(\rho^2) &=& \lambda^{-2/3} H^{4/3} M_p^2 \times 2 (\Delta y) f^\prime(\rho^2)  \nonumber \\
\partial_a f(\rho^2) &=& \lambda^{4/3} H^{4/3} M_p^2 \times 2 (\Delta z) f^\prime(\rho^2)
 \nd
For a power-law function $f$, which is typical of these ``lump" solutions, we have

\bg
f^\prime(\rho^2) \propto \rho^{-2} f(\rho^2)
\nd 
and the $\rho^{-2}$ will cancel the extra scaling factors in \eqref{dperp}, so that we have $\partial_{\perp} f(\rho^2) \sim f(\rho^2)$ in terms of scaling. This means that 

\bg
\partial_{y_\perp} G^{(loc)} \sim \partial_{y_\|} G^{(loc)}
\nd
in terms of their contribution to the scaling after complete contraction. However, $\partial_{a} G^{(loc)}$ also no longer vanishes and does not scale the same way as $\partial_y G^{(loc)}$, but rather as

\bg
\partial_a G^{(loc)} \sim \lambda^{-1/3} H^{-4/3} M_p^{-2}
\nd

The flux equation of motion still requires that $G^{(loc)}_{mnpa} \sim H^2$ just as for the bulk fluxes. In fact the scaling contribution from $G_{(loc)}$ is exactly the same as from $G_{(int)}$, but derivatives of these fluxes along the orthogonal directions now allow for combinations such as:

\bg
\partial_{a} G^{(loc)} \partial_{b} G^{(loc)} \partial_{n} G^{(loc)}  \sim \lambda^0 H^{-4} M_p^{-6}
\nd

The $\lambda^0$ scaling, makes these combinations precisely example of the ``non-scaling operators" described in section \ref{dsp}. Contractions involving these operators allows us to generate an infinite family of higher derivative corrections that have the same $\lambda$-scalings. One has to be careful, however, as many of these terms may vanish on-shell, as their form is constrained by the flux equations of motion and bianchi-identity.

A similar thing happens with the Riemann tensor in the localized regions. Since we assumed that $h_{mn}$ scales the same as $\tilde{g}_{mn}$ there is no difference in terms of scaling between raising indices with $g^{mn}$ or $\tilde{g}^{mn}$. This means that in the expression for the Riemann tensor, the inverse metric scalings will still cancel with the metric scalings. However, there are now new non-vanishing contributions to the Riemann tensor, containing $z^a$ derivatives.

\bg
R_{(loc)} = R_{(yy)} + R_{(yz)} + R_{(zz)}
\nd
where the subscripts denote the subspaces along which the derivatives inside the Riemann tensors act.

Each of these terms will have a different scaling and must therefore vanish on its own or cancel against other terms in the EOM with the same scaling. Specifically we have $R_{(yy)}$ still having the same scaling as the non-localized curvature, while

\bg
R_{(yz)} \sim \lambda^{-1/3} H^{-4/3} M_p^{-2}   ~~~~~~~~
R_{(zz)} \sim \lambda^{-4/3} H^{-4/3} M_p^{-2}.
\nd

Note that these two components have a completely distinct scaling from all the other terms and must therefore satisfy the equations of motion independently. This may be understood as the statement that the localized geometry dual to the D7 branes and O7 planes must be a solution of the equations of motion in its own right. These new curvature components can then form new higher-derivative corrections, of the form

\bg
R_{(zz)} R_{(yy)} R_{(yy)}  \sim \lambda^0 H^{-4} M_p^{-6},
\nd
which notably contribute the same scalings as the operators build out of flux gradients.

An important detail is that these non-scaling terms that we have found do not seem related to the expansion of the DBI action. The different nature of our non-scaling terms can be seen by observing that the terms in the DBI action are related to the derivative expansion along the worldvolume directions, whereas both $R_{(loc)}$ and $\nabla_\perp G_{(loc)}$ contain orthogonal derivatives.  Instead, we interpret these terms as being related to string loop corrections to the worldvolume action on the IIB side. Indeed, the presence of derivatives along the normal directions in these terms appears to probe the effective thickness of the branes, which is related to the string coupling.

As discussed in section \ref{dsp}, even though these terms all have the same $\lambda$ scaling, they are still distinguished by their scaling with respect to the other parameters, or more specifically, by the magnitudes of $R_{(loc)}$ and $G_{(loc)}$ relative to $H^{4/3} M_p^2$. When these quantities are small, this provides additional expansion parameters that suppress the higher-order contractions of these terms and allow for a truncation of the EOM. However, if the solution we are seeking requires strong worldvolume fluxes or curvatures, we will obtain a breakdown of the effective field theory description, as all the members of this family of corrections will enter on equal footing.

Recall that our goal is to cancel the terms in the equations of motion coming from $R_{(ext)}$, which is related to the the 4-dimensional Einstein tensor on the IIB side. One approach we could have imagined, was constructing contractions of the 4-dimensional Riemann tensor with some perturbative corrections. The $\lambda^0$ corrections already respect the matching of their time-depenences. However the localized nature of these corrections, means we can not use them to cancel $R_{(ext)}$ everywhere along the internal manifold. Furthermore, in order to cancel the classical term, the additional factors would have to be of order one, thus leaving the weak flux/curvature limit and leading to a breakdown of the EFT description.

What seems to be required is that we construct quantum corrections out of the \emph{other} two-derivative terms, contracted with something that will shift their scaling to match that of $R_{(ext)}$. In the next section we will analyze the non-local terms and find precisely such a term, which will, perhaps unsurprisingly, turn out to be related to brane-instantons wrapping 4-cycles of the internal manifold.

\subsection{Non-local terms and non-perturbative corrections}

In this section we study generic non-local effects (considered in detail in \cite{ME2}) given by integral expressions of the form

\bg \label{npGenTerm}
S^{(nloc)} = M_p^{22}\int  d^{11} x d^{11} x^\prime \sqrt{g(x)~ g(x^\prime) }  O(x) G(x - x_{(1)}) O^\prime(x^\prime) 
\nd

where $O$ and $O^\prime$ are some completely contracted products of fields and $G(x-x^\prime)$ is some appropriate non-locality function, which depends on the correction.

If we take $G$ be a normalized delta function, it eats up the integral and we recover the local term $(O O^\prime)(x) $,\footnote{Unless the operator $O^\prime$ is itself further non-local, which we might also wish to allow.} but if it is delocalized along some directions, we obtain contributions that represent configurations of extended objects, including spacelike configurations, such as brane-instantons. The exact form of the allowed corrections will of course be constrained by the various symmetries of the theory. Unless the operator $O^\prime$ contains sufficiently many derivatives, the overall power of $M_p$ will be positive relative to the classical terms. This would make the term appear to dominate in the low-energy limit. The way out of this conundrum is to recognize that the positive power of $M_p$ indicates that the term is non-perturbative in $M_p^{-1}$ and the usual exponentiation of non-perturbative corrections via the dilute gas approximation should take place. This means that these corrections actually appear in the action as non-perturbative exponential terms of the form

\bg \label{nonLoc}
S^{(nloc)} &=& M_p^{11} \int d^{11} x \sqrt{g(x)}~ O(x) \left( a_1 e^{-\mathcal{I}(x)} + a_2 e^{-2\mathcal{I}(x)} + ... \right) \\
\mathcal{I}(x) &=& \int d^{11} x^\prime \sqrt{g(x^\prime)} G(x-x^\prime) O^\prime(x^\prime) \nonumber
\nd
which has the desired low-energy behavior. This, of course, gives our action precisely the trans-series structure that we discussed in section \ref{EFT}. Note that in doing the dilute gas calculation, we have performed the path integral the over the transverse positions of the extended object, so even though the object itself is partially localized, the exponential term can be evaluated at any point and will capture the properly weighed contribution of the configuration corresponding to the extended object overlapping that location.

The stress tensor coming from the leading non-perturbative contribution is then given by

\bg \label{nonLocT}
T^M_{~~N}(x) = g^{MP} \left(   \left( \frac{\delta O(x) }{\delta g^{PN}(x)} - g_{PN} O(x) \right) e^{-\mathcal{I}(x)} -  \int d^{11}x^\prime \sqrt{g} ~ O(x^\prime) \frac{\delta \mathcal{I}(x^\prime)}{\delta g^{PN}(x) }~ e^{-\mathcal{I}(x^\prime)} \right) , \nonumber \\
\nd
We see that both terms carry the non-perturbative exponential suppression $e^{-\mathcal{I} }$. Besides this non-perturbative scaling, the first term has the same scaling as $O(x)$ itself, while the second term has the same scaling as $O(x) \mathcal{I}(x)$, which will be lower order in $M_p^{-1}$, since $\mathcal{I}(x)$ has positive $M_p$ scaling. Thus, the second term will give a new contribution to the equations of motion, which will also be the dominant contribution, while the first term will simply give a non-perturbativelly suppressed correction to the already existing stress tensor coming from $O(x)$. The overall scaling of this operator depends on the scalings of all the operators that enter $\mathcal{I}$ as well as the scalings of the functions $G$. Here we will make a conservative assumption that the functions $G$ have no inherent scaling of their own.

Let us consider a particular example of such a correction of the form

\bg \label{M5inst}
\mathcal{I}(x^\prime) = M_p^{11} \int d^4 y_\| d^2 z \sqrt{g_6} \int d^3 x d^2 y_\perp \sqrt{g_3 g_2} \delta^{(3)}(x-x^\prime) \delta^{(2)}(y_\perp - y^\prime_\perp) ,
\nd

which we obtain from \eqref{npGenTerm} by choosing $O^\prime = 1$ and the function $G$ to be extended along a 4-cycle of the internal manifold and the torus, parametrized by $y_\|$ and $z$ respectively, and localized in the other directions. This integral is related to the worldvolume term in the action of an M5 instanton, which is dual to the D3 instantons used to generate the non-perturbative contributions in KKLT and similar scenarios.\footnote{If instead of the torus, it wraps the non-trivial 2-cycles appearing in the multi-centered lump solutions describing our D7/O7 stacks, they become stuck at the lump locus and correspond to a worldvolume instanton density related to worldvolume gauge theory effects like gaugino condensation.}

The choice of this particular correction is guided by that fact that it scales as 

\bg
\mathcal{I}(x^\prime) \sim \lambda^0 H^4 M_p^6,
\nd

which in particular means that it's independent of the external spacetime coordinates and will neither grow nor vanish in the $\lambda \to 0 $ limit, where we expect to recover the type IIB description. If we had chosen to fewer torus directions or more of any other directions, the resulting non-local term would have negative $\lambda$ scaling and the corresponding non-perturbative exponential would vanish in that limit. As an aside, the only other $\lambda^0$-scaling contribution can come from wrapping one torus direction and two internal directions. We recognize this as an M2-instanton, which is dual to type IIB worldsheet or D1- instantons wrapping 2-cycles on the internal manifold.

For the M5-instanton case, we can now consider a non-perturbative correction of the form

\bg \label{npcorrect}
\int d^{11} x \sqrt{g} O(x) e^{-\mathcal{I}(x)} 
\nd
where $O(x)$ is any of the 2-derivative operators in the bulk action other than $R_{(ext)}$. Note that  
\bg \label{Ivar}
\frac{ \delta \mathcal{I}(x^\prime)}{\delta g^{MN}(x) } = - g_{MN}(x)
\nd
and so the interesting new contribution to the stress tensor (i.e. the second term in \eqref{nonLocT}) will be

\bg
T^{(nloc)M}_{~~~~~~~N}(x) = g^M_{~~N} (x) \int d^{11}x^\prime \sqrt{g_6} O(x^\prime) e^{-\mathcal{I}(x^\prime)}
\nd
 The scaling of this contribution is

\bg
T^{(nloc)M}_{~~~~~~~N}(x) \sim \lambda^{2/3} H^{8/3} M_p^{-2} \times (M_p^6 e^{-H^4 M_p^6} )
\nd

The factor in the parentheses is an exponentially small time-independent term, while the pre-factor has exactly the correct scaling as $R_{(ext)}$. Keeping track of the signs, we can see that it can precisely cancel the scaling of $R_{(ext)}$ when

\bg
\Lambda = -M_p^6 e^{-\langle H^4\rangle M_p^6} 
\nd

The mass-dimensions of this expression may look off, but recall that $\Lambda$ is the 4-dimensional cosmological constant on the IIB side, while $M_p$ is the 11-dimensional Planck mass. The conversion factors between them contain powers of the planck mass. We can work out the conversions to write the result in perhaps a more illuminating form:

\bg
\Lambda \langle H^4 \rangle \sim \left(\frac{m_{(4)}}{M_{(IIB)}}\right)^2 e^{-\left(\frac{m_{(4)}}{M_{(IIB)}}\right)^2}
\nd

where $M_{(IIB)}$ is the 10-dimensional type IIB Planck mass, and we omitted possible numerical coefficients both in the prefactor and the exponent. Recall that small $\Lambda \langle H^4 \rangle$ indicates large scale separation, exactly as desired.

Thus it appears that by including the non-perturbative effects of M5 instantons, we very readily obtain a scale separated type IIB AdS solution. We recognize this as essentially the same as the KKLT AdS vacuum. In our approach, however, we can also see that using these non-perturbative effects was really the only option from the start. The perturbative terms simply do not have the correct scaling to cancel the EOM-violating terms coming from the 4-dimensional curvature. This may raise questions regarding alternative mechanisms of Kahler moduli stabilization, particularly those that only use perturbative effects, but we will not pursue this issue here.

Note that the large scale-separation ultimately stems from the non-perturbative nature of the correction responsible for it and the fact that it exponentiates due to the standard dilute gas argument. It is noteworthy that our motivation to exponentiate the non-perturbative terms comes from requiring the correct behavior as $M_p \to \infty$. The fact that these terms are also non-perturbative in $H^{-1}$ and are related to the Kahler moduli of the internal manifold appears as a happy coincidence. However, we could have also arrived at this idea by recognizing that to obtain the scaling of $R_{(ext)}$ from the other terms, we needed to contract them with something that had positive $H$ scaling. Since our EFT regime is defined by expanding in inverse powers of $H$, the necessary factor would have to be non-perturbative in $H$ as well. 

The fact that these factors also end up $\lambda$-independent is a further non-trivial feature. Note that corrections coming from M2-instantons, while also $\lambda$-independent, would not give the required $H^4$ scaling. We have therefore identified M5-instantons (and therefore D3-instantons, or their fractional versions in the form of worldvolume gaugino condensation on the IIB side) as essentially the unique type of correction that allows for non-zero curvature in the external directions, in scale separated compactifications without smeared sources.

This uniqueness, however, appears to present an obstacle for obtaining de Sitter solutions, as this would require us to flip the sign of $T^{(nloc)}$. Note that we are free to dress our non-local correction with any of the non-scaling terms that we have described in section \ref{locterm}, so it's conceivable that, say, by placing appropriate worldvolume fluxes on the D7-branes, we can generate terms of the form \eqref{npcorrect} with $O$ dressed with some non-scaling operators which flip the sign of the contribution. This is essentially the scenario of \cite{BKQ}, which is similar to KKLT except the anti-brane charge comes from SUSY-breaking worldvolume fluxes.

Note, however that our analysis suggests that it will not be the classical terms coming from these fluxes that are responsible for the uplift (which is consistent with the classical de Sitter no-go theorems), nor from the higher order terms in the DBI action, but rather from higher loop corrections that contain powers of these fluxes. However, since the original non-dressed contribution is still present, the non-scaling factors would have to be of order one, just to cancel the original negative contribution to the cosmological constant. In other words, the equation of motion would contain a part that schematically looks like

\bg
\frac{\delta}{\delta g^{MN} } \sqrt{g} \left(  R_{(ext)} + O_{c}(x) e^{-\mathcal{I}(x)} + O_{q}O_{c}(x) e^{-\mathcal{I}(x)}  \right) = 0
\nd

Where $O_c$ are the bulk classical 2-derivative terms, and $O_q$ are the localized perturbative factors made up of $\nabla_\perp G$ and $R_{(loc)}$ described in section \ref{locterm}. In order to reverse the effect of the second term, $O_q$ needs to have magnitude of order one. However, this would destroy the relative suppression between the different operators $O_q$. This would lead to the entire infinite family of non-scaling operators contributing to the equations of motion and lead to a breakdown of the effective field theory description, precisely along the lines described in section \ref{dsp}.

\section{Summary and Discussion} \label{conclusion}

In this paper, we have adopted the approach of viewing effective theories as truncations of a resurgent trans-series, which represents the full non-perturbative quantum effective action. When the full theory has no dimensionless parameters, as is the case in string and M-theory, a given background ansatz fixes the scalings with respect to all the couplings and mass scales, as those arise from expectation values of fields. This allows us to split the various terms in the full quantum equations of motion according to their scalings, which in turn allows us to determine whether the equation of motion may be truncated, and therefore whether a given putative solution resides within the EFT regime that defines the trans-series expansion in the first place.

We applied this approach to a type IIB scale-separated (A)dS ansatz, by expanding around the large-volume, low energy regime, which we studied in terms of M-theory fields for convenience, and found no obstructions to scale separated AdS solutions. Furthermore, the ingredients required to construct these solutions are essentially uniquely determined to be those used in the KKLT and similar scenarios, i.e. brane-instantons wrapping 4-cycles in the internal manifold. De Sitter solutions require changing the sign of the non-perturbative contribution, by dressing these non-perturbative contributions with operators that have no scaling with respect to the parameters we studied. We have identified such operators, and argued that they are related to loop corrections in the D7 worldvolume theory. These corrections form an infinite series which is controlled at weak worldvolume flux and extrinsic curvature, however our analysis indicates that lifting to de Sitter space requires that these fields are not weak and thus leads to a breakdown of the EFT description, as infinitely many unsuppressed contributions to the equations of motion appear.

For Anti-de Sitter space, our results appear in conflict with the proposed AdS swampland conjectures, which state that AdS compactifications can not have large scale separation. It is worth remembering that these conjectures are motivated by studying the flat space limit of non-scale-separated Freund-Rubin type AdS solutions, which solve the 2-derivative EOM. The solutions we studied here do not belong to this family, and indeed do not even solve the 2-derivative EOM, but require non-perturbative effects to be included. 

Thus it appears that a more careful statement would perhaps be that scale-separated and non-scale-separated AdS solutions lie far from each other in configuration space and are not part of the same EFT. Thus if one starts in the Freund-Rubin-like regime, one can not reach the scale-separated solutions without leaving the regime of validity of that EFT. Similarly, if we start in our scale-separated regime, we could not reach the the non-scale-separated regime either. Moving between the two-regimes, should then be described by a duality transformation, rather than a motion within the field space of a single EFT. However, once we are in the scale-separated regime, our approach does not seem to indicate any fundamental obstacles to having AdS solutions and indeed shows us precisely which ingredients are required to make it work. 

Note that our use of non-perturbative effects circumvents the no-go theorem of \cite{AdSnogo}, which is formulated at the two-derivative level. On the other hand, although the examples we consider contain orientifold planes, which also circumvent the no-go theorem, we make no explicit use of their properties so it is possible that scale-separated AdS compactifications are possible if not in the absense of orientifolding, then at least without them being crucial in generating the non-zero spacetime curvature.

As for de Sitter compactifications, the results presented here do not necessarily invalidate the existence of time-independent de Sitter vacua {\it per se}, although their non-existence would certainly explain our results as well. Putting aside this pessimistic scenario, our results do, however, put into question whether it is appropriate to view de Sitter vacua as states within the same EFT as the SUSY compactifications that serve as starting points for dS constructions. If the answer is negative, then a duality transformation is necessary to relate dS compactifications to Minkowski or scale-separated AdS compactifications.

This is in contrast to the scenario studied in \cite{ME2} where time-dependence was switched on for the internal fluxes and compactification manifold, shifting the $\lambda$ scalings of the terms and restoring their relative suppression. Doing so appears to open the door to solutions whose late-time behavior is de Sitter which retain an EFT description purely in terms of supergravity with non-trivial but controlled quantum corrections.

We can try to relate our rather general results to some more specific questions in the context of KKLT and similar constructions which rely on uplifting a scale-separated AdS vacuum to a dS vacuum.\footnote{Our ansatz is different in certain details from the KKLT scenario, in that we have essentially dissolved the anti-D3 charge in the worldvolume fluxes. In this respect our scenario is more similar to \cite{BKQ}. However, since our analysis relies primarily on the scalings of the various terms rather than their exact physical origin, we expect the general lessons to be similar.} Having constructed the scale-separated AdS vacuum, we face the task of uplifting. There are two questions worth separating here: i) does the uplift procedure actually uplift past zero curvature to de Sitter space and ii) does it ruin the EFT description in doing so?

Our present analysis doesn't have much to say about the first question, as the answer depends on the specific dynamics within the KKLT scenario. The latter question boils down to the whether introducing anti-branes or similar ingredients to an AdS compactification can be regarded as a motion in field space of the original AdS EFT, without ever leaving its regime of validity. This, in turn, is related to the question of whether the KPV saddle \cite{KPV} can be regarded as a state in the Klebanov-Strassler field theory \cite{KlebanovStrassler}, or whether the brane-flux annihilation process that tunnels between them involves effects outside the EFT description and the states therefore lie in different EFTs. \footnote{Note that this problem is independent of whether or not the KPV state actually yields a de Sitter geometry upon compactification.}

If we wish to have a 4-dimensional effective theory, we must use a cutoff for our EFT that excludes KK modes. In this case, the results of section \ref{metric} indicate that the answer to the above questions is in the negative and that the uplift procedure takes us outside the regime of the effective theory where the AdS vacuum resides. Thus, if time-independent de Sitter compactifications exist, the dynamics around these states would be described by some other EFT that is related to the EFTs around SUSY compactifications by the introduction of new degrees of freedom and a duality transformation, rather than a simple shift in the EFT field-space. This conclusion is consistent with, for example, the results of \cite{Aalsma}, where the inclusion of KK modes was argued to be necessary to describe both the KPV and the SUSY vacuum as states in the same theory. This is also consistent with the more recent formulations of the KKLT construction in the presence of anti-branes in terms of ``de Sitter supergravity'' with constrained superfields as described, for example, in \cite{dSSUGRA}. It is possible that the passage from the usual 4D supergravity regime (with quantum corrections) to ``de Sitter supergravity'' can also be regarded as a duality transformation.

\section*{Acknowledgements}

The author would like to thank Keshav Dasgupta for helpful discussions and feedback on the draft.

{}

\begin{thebibliography}{}

\bibitem{vafa1}
  C.~Vafa,
  ``The String landscape and the swampland,''
  hep-th/0509212.

\bibitem{swamplandreview}
  T.~D.~Brennan, F.~Carta and C.~Vafa,
  ``The String Landscape, the Swampland, and the Missing Corner,''
  PoS TASI {\bf 2017}, 015 (2017)
  %doi:10.22323/1.305.0015
  [arXiv:1711.00864 [hep-th]].

 E.~Palti,
  ``The Swampland: Introduction and Review,''
  Fortsch.\ Phys.\  {\bf 67}, no. 6, 1900037 (2019)
  %doi:10.1002/prop.201900037
  [arXiv:1903.06239 [hep-th]].

\bibitem{vafa-dS}
  G.~Obied, H.~Ooguri, L.~Spodyneiko and C.~Vafa,
  ``De Sitter Space and the Swampland,''
  arXiv:1806.08362 [hep-th];

\bibitem{Krishnan}
  S.~K.~Garg and C.~Krishnan,
  ``Bounds on Slow Roll and the de Sitter Swampland,''
  arXiv:1807.05193 [hep-th];


\bibitem{dSrefine}
H.~Ooguri, E.~Palti, G.~Shiu and C.~Vafa,
  ``Distance and de Sitter Conjectures on the Swampland,''
  Phys.\ Lett.\ B {\bf 788}, 180 (2019)
  %doi:10.1016/j.physletb.2018.11.018
  [arXiv:1810.05506 [hep-th]].


\bibitem{noEI}

  T.~Rudelius,
  ``Conditions for (No) Eternal Inflation,''
  JCAP {\bf 1908}, 009 (2019)
  %doi:10.1088/1475-7516/2019/08/009
  [arXiv:1905.05198 [hep-th]].

\bibitem{TCC}
  A.~Bedroya and C.~Vafa,
  `Trans-Planckian Censorship and the Swampland,''
  arXiv:1909.11063 [hep-th].

  A.~Bedroya, R.~Brandenberger, M.~Loverde and C.~Vafa,
  ``Trans-Planckian Censorship and Inflationary Cosmology,''
  arXiv:1909.11106 [hep-th].

\bibitem{AdSdist}
  D.~Lüst, E.~Palti and C.~Vafa,
  ``AdS and the Swampland,''
  Phys.\ Lett.\ B {\bf 797}, 134867 (2019)
  %doi:10.1016/j.physletb.2019.134867
  [arXiv:1906.05225 [hep-th]].




\bibitem{AdSnogo}
  F.~F.~Gautason, M.~Schillo, T.~Van Riet and M.~Williams,
  ``Remarks on scale separation in flux vacua,''
  JHEP {\bf 1603}, 061 (2016)
  %doi:10.1007/JHEP03(2016)061
  [arXiv:1512.00457 [hep-th]].


\bibitem{KKLT}
  S.~Kachru, R.~Kallosh, A.~D.~Linde and S.~P.~Trivedi,
  ``De Sitter vacua in string theory,''
  Phys.\ Rev.\ D {\bf 68}, 046005 (2003)
  %doi:10.1103/PhysRevD.68.046005
  [hep-th/0301240].
  
  \bibitem{BKQ} 
  C.~P.~Burgess, R.~Kallosh and F.~Quevedo,
  ``De Sitter string vacua from supersymmetric D terms,''
  JHEP {\bf 0310}, 056 (2003)
  %doi:10.1088/1126-6708/2003/10/056
  [hep-th/0309187].

\bibitem{LVS}

  V.~Balasubramanian, P.~Berglund, J.~P.~Conlon and F.~Quevedo,
  ``Systematics of moduli stabilisation in Calabi-Yau flux compactifications,''
  JHEP {\bf 0503}, 007 (2005)
  %doi:10.1088/1126-6708/2005/03/007
  [hep-th/0502058].

  A.~Westphal,
  ``de Sitter string vacua from Kahler uplifting,''
  JHEP {\bf 0703}, 102 (2007)
  %doi:10.1088/1126-6708/2007/03/102
  [hep-th/0611332].

\bibitem{StarWars1}
  R.~Kallosh, F.~Quevedo and A.~M.~Uranga,
  ``String Theory Realizations of the Nilpotent Goldstino,''
  JHEP {\bf 1512}, 039 (2015)
  %doi:10.1007/JHEP12(2015)039
  [arXiv:1507.07556 [hep-th]].

E.~A.~Bergshoeff, K.~Dasgupta, R.~Kallosh, A.~Van Proeyen and T.~Wrase,
  ``$ \overline{\mathrm{D}3} $ and dS,''
  JHEP {\bf 1505}, 058 (2015)
    [arXiv:1502.07627 [hep-th]];

       R.~Kallosh, A.~Linde, E.~McDonough and M.~Scalisi,
  ``de Sitter Vacua with a Nilpotent Superfield,''
  Fortsch.\ Phys.\  {\bf 67}, no. 1-2, 1800068 (2019)
   [arXiv:1808.09428 [hep-th]]; ``4D models of de Sitter uplift,''
  Phys.\ Rev.\ D {\bf 99}, no. 4, 046006 (2019)
    [arXiv:1809.09018 [hep-th]]; ``dS Vacua and the Swampland,''
  JHEP {\bf 1903}, 134 (2019)
    [arXiv:1901.02022 [hep-th]];

Y.~Akrami, R.~Kallosh, A.~Linde and V.~Vardanyan,
``The Landscape, the Swampland and the Era of Precision Cosmology,''
Fortsch. Phys. \textbf{67}, no.1-2, 1800075 (2019)
%doi:10.1002/prop.201800075
[arXiv:1808.09440 [hep-th]].

Y.~Hamada, A.~Hebecker, G.~Shiu and P.~Soler,
``On brane gaugino condensates in 10d,''
JHEP \textbf{04}, 008 (2019)
%doi:10.1007/JHEP04(2019)008
[arXiv:1812.06097 [hep-th]].

  R.~Kallosh,
  ``Gaugino Condensation and Geometry of the Perfect Square,''
  Phys.\ Rev.\ D {\bf 99}, no. 6, 066003 (2019)
  %doi:10.1103/PhysRevD.99.066003
  [arXiv:1901.02023 [hep-th]].

F.~Carta, J.~Moritz and A.~Westphal,
``Gaugino condensation and small uplifts in KKLT,''
JHEP \textbf{08}, 141 (2019)
%doi:10.1007/JHEP08(2019)141
[arXiv:1902.01412 [hep-th]].

Y.~Hamada, A.~Hebecker, G.~Shiu and P.~Soler,
``Understanding KKLT from a 10d perspective,''
JHEP \textbf{06}, 019 (2019)
%doi:10.1007/JHEP06(2019)019
[arXiv:1902.01410 [hep-th]].

\bibitem{StarWars2}
  U.~H.~Danielsson and T.~Van Riet,
  ``What if string theory has no de Sitter vacua?,''
  Int.\ J.\ Mod.\ Phys.\ D {\bf 27}, no. 12, 1830007 (2018)
  %doi:10.1142/S0218271818300070
  [arXiv:1804.01120 [hep-th]].

  J.~Moritz, A.~Retolaza and A.~Westphal,
  ``Toward de Sitter space from ten dimensions,''
  Phys.\ Rev.\ D {\bf 97}, no. 4, 046010 (2018)
  %doi:10.1103/PhysRevD.97.046010
  [arXiv:1707.08678 [hep-th]].


  F.~F.~Gautason, V.~Van Hemelryck and T.~Van Riet,
  ``The Tension between 10D Supergravity and dS Uplifts,''
  Fortsch.\ Phys.\  {\bf 67}, no. 1-2, 1800091 (2019)
  %doi:10.1002/prop.201800091
  [arXiv:1810.08518 [hep-th]].

  U.~Danielsson,
  ``The quantum swampland,''
  JHEP {\bf 1904}, 095 (2019)
%  doi:10.1007/JHEP04(2019)095
  [arXiv:1809.04512 [hep-th]].

  F.~F.~Gautason, V.~Van Hemelryck, T.~Van Riet and G.~Venken,
  ``A 10d view on the KKLT AdS vacuum and uplifting,''
  arXiv:1902.01415 [hep-th].

  S.~Sethi,
  ``Supersymmetry Breaking by Fluxes,''
  JHEP {\bf 1810}, 022 (2018)
  %doi:10.1007/JHEP10(2018)022
  [arXiv:1709.03554 [hep-th]].

  I.~Bena, E.~Dudas, M.~Graña and S.~Lüst,
``Uplifting Runaways,''
  Fortsch.\ Phys.\  {\bf 67}, no. 1-2, 1800100 (2019)
  [Fortsch.\ Phys.\  {\bf 2018}, 1800100]
%  doi:10.1002/prop.201800100
  [arXiv:1809.06861 [hep-th]].

\bibitem{ME0}

  K.~Dasgupta, R.~Gwyn, E.~McDonough, M.~Mia and R.~Tatar,
  ``de Sitter Vacua in Type IIB String Theory: Classical Solutions and Quantum Corrections,''
  JHEP {\bf 1407}, 054 (2014)
  %doi:10.1007/JHEP07(2014)054
  [arXiv:1402.5112 [hep-th]].

\bibitem{ME}

  K.~Dasgupta, M.~Emelin, E.~McDonough and R.~Tatar,
  ``Quantum Corrections and the de Sitter Swampland Conjecture,''
  JHEP {\bf 1901}, 145 (2019)
  %doi:10.1007/JHEP01(2019)145
  [arXiv:1808.07498 [hep-th]].

\bibitem{ME2}

K.~Dasgupta, M.~Emelin, M.~M.~Faruk and R.~Tatar,
  ``de Sitter Vacua in the String Landscape,''
  arXiv:1908.05288 [hep-th].

 \bibitem{ME3}
  K.~Dasgupta, M.~Emelin, M.~M.~Faruk and R.~Tatar,
 ``How a four-dimensional de Sitter solution remains outside the swampland,''
  arXiv:1911.02604 [hep-th].




\bibitem{transseries}
 D.~Dorigoni,
  ``An Introduction to Resurgence, Trans-Series and Alien Calculus,''
  Annals Phys.\  {\bf 409}, 167914 (2019)
  %doi:10.1016/j.aop.2019.167914
  [arXiv:1411.3585 [hep-th]].

G.~V.~Dunne and M.~Ünsal,
  ``What is QFT? Resurgent trans-series, Lefschetz thimbles, and new exact saddles,''
  PoS LATTICE {\bf 2015}, 010 (2016)
  %doi:10.22323/1.251.0010
  [arXiv:1511.05977 [hep-lat]].
  

I.~Aniceto, G.~Basar and R.~Schiappa,
``A Primer on Resurgent Transseries and Their Asymptotics,''
Phys. Rept. \textbf{809}, 1-135 (2019)
%doi:10.1016/j.physrep.2019.02.003
[arXiv:1802.10441 [hep-th]].


\bibitem{DS}
  M.~Dine and N.~Seiberg,
  ``Is the Superstring Weakly Coupled?,''
  Phys.\ Lett.\  {\bf 162B}, 299 (1985).
  
\bibitem{DRS}
   K.~Dasgupta, G.~Rajesh and S.~Sethi,
  ``M theory, orientifolds and G - flux,''
  JHEP {\bf 9908}, 023 (1999)
  %doi:10.1088/1126-6708/1999/08/023
  [hep-th/9908088].
  
  
\bibitem{GKP}
S.~B.~Giddings, S.~Kachru and J.~Polchinski,
  ``Hierarchies from fluxes in string compactifications,''
  Phys.\ Rev.\ D {\bf 66}, 106006 (2002)
  %doi:10.1103/PhysRevD.66.106006
  [hep-th/0105097].


\bibitem{fluxAdS}
 D.~Lust and D.~Tsimpis,
  ``Classes of AdS(4) type IIA/IIB compactifications with SU(3) x SU(3) structure,''
  JHEP {\bf 0904}, 111 (2009)
  %doi:10.1088/1126-6708/2009/04/111
  [arXiv:0901.4474 [hep-th]].

\bibitem{MNnogo}
  J.~M.~Maldacena and C.~Nunez,
 ``Supergravity description of field theories on curved manifolds and a no go theorem,''
  Int.\ J.\ Mod.\ Phys.\ A {\bf 16}, 822 (2001)
 % doi:10.1142/S0217751X01003935, 10.1142/S0217751X01003937
  [hep-th/0007018].

\bibitem{sen}
A.~Sen,
  ``F theory and orientifolds,''
  Nucl.\ Phys.\ B {\bf 475}, 562 (1996)
  %doi:10.1016/0550-3213(96)00347-1
  [hep-th/9605150].

\bibitem{KS}
  K.~Dasgupta and S.~Mukhi,
  ``F theory at constant coupling,''
  Phys.\ Lett.\ B {\bf 385}, 125 (1996)
  %doi:10.1016/0370-2693(96)00875-1
  [hep-th/9606044].
  
  \bibitem{lumps} 
  T.~Maxfield and S.~Sethi,
  ``DBI from Gravity,''
  JHEP {\bf 1702}, 108 (2017)
  %doi:10.1007/JHEP02(2017)108
  [arXiv:1612.00427 [hep-th]].
  
  \bibitem{KPV}
   S.~Kachru, J.~Pearson and H.~L.~Verlinde,
  ``Brane / flux annihilation and the string dual of a nonsupersymmetric field theory,''
  JHEP {\bf 0206}, 021 (2002)
  %doi:10.1088/1126-6708/2002/06/021
  [hep-th/0112197].
  
  \bibitem{KlebanovStrassler}
    I.~R.~Klebanov and M.~J.~Strassler,
  ``Supergravity and a confining gauge theory: Duality cascades and chi SB resolution of naked singularities,''
  JHEP {\bf 0008}, 052 (2000)
  %doi:10.1088/1126-6708/2000/08/052
  [hep-th/0007191].
  
  \bibitem{Aalsma}
  
   L.~Aalsma, J.~P.~van der Schaar and B.~Vercnocke,
  ``Constrained superfields on metastable anti-D3-branes,''
  JHEP {\bf 1705}, 089 (2017)
  %doi:10.1007/JHEP05(2017)089
  [arXiv:1703.05771 [hep-th]]
  
    L.~Aalsma, M.~Tournoy, J.~P.~Van Der Schaar and B.~Vercnocke,
  ``Supersymmetric embedding of antibrane polarization,''
  Phys.\ Rev.\ D {\bf 98}, no. 8, 086019 (2018)
  %doi:10.1103/PhysRevD.98.086019
  [arXiv:1807.03303 [hep-th]].
  
  \bibitem{dSSUGRA}
  E.~A.~Bergshoeff, D.~Z.~Freedman, R.~Kallosh and A.~Van Proeyen,
  ``Pure de Sitter Supergravity,''
  Phys.\ Rev.\ D {\bf 92}, no. 8, 085040 (2015)
  Erratum: [Phys.\ Rev.\ D {\bf 93}, no. 6, 069901 (2016)]
  %doi:10.1103/PhysRevD.93.069901, 10.1103/PhysRevD.92.085040
  [arXiv:1507.08264 [hep-th]].
  
\end{thebibliography}
 \end{document}